\documentclass[aps,showpacs,prb,superscriptaddress,floatfix,twocolumn,citeautoscript]{revtex4-1}
\usepackage{graphicx}
\usepackage{amsmath}
\usepackage{amssymb}
\usepackage{bm}
\usepackage{dcolumn}
\usepackage{subfigure} 
\usepackage{hyperref}
\usepackage[usenames]{color}
\usepackage{booktabs}
\usepackage{multirow}
\usepackage{siunitx}
\usepackage{wrapfig}
\usepackage{lipsum}
\usepackage{setspace}
\hypersetup{pdfborder=0 0 0,colorlinks=true,citecolor=blue,linkcolor=blue}
\input epsf
\newcommand{\BN}{\textit{h}-BN}

\begin{document}

\title{Anisotropic hybrid excitation modes in monolayer and double-layer phosphorene on polar substrates} 
\author{S. Saberi-Pouya}
\affiliation{Department of Physics, Shahid Beheshti University, G.C., Evin, Tehran 1983969411, Iran}
\affiliation{Department of Physics, University of Antwerp, Groenenborgerlaan 171, B-2020 Antwerpen, Belgium}
\author{T. Vazifehshenas}
\email{t-vazifeh@sbu.ac.ir}
\affiliation{Department of Physics, Shahid Beheshti University, G.C., Evin, Tehran 1983969411, Iran}
\author{T. Salavati-fard}
\affiliation{Department of Physics and Astronomy, University of Delaware, Newark, DE 19716, USA}
\author{M. Farmanbar}
\affiliation{Faculty of Science and Technology and MESA+ Institute for Nanotechnology, University of Twente, P.O. Box 217, 7500 AE Enschede, The Netherlands}

\begin{abstract}
	
We investigate the anisotropic hybrid plasmon-SO phonon dispersion relations in monolayer and double-layer phosphorene systems located on the polar substrates, such as SiO$_{2}$, \BN\ and Al$_{2}$O$_{3}$. We calculate these hybrid modes with using the dynamical dielectric function in the RPA by considering the electron-electron interaction and long-range electric field generated by the substrate SO phonons via Fr$\ddot{\mathrm{o}}$hlich interaction. In the long-wavelength limit, we obtain some analytical expressions for the hybrid plasmon-SO phonon dispersion relations which represent the behavior of these modes akin to the modes obtaining from the loss function. Our results indicate a strong anisotropy in plasmon-SO phonon modes, whereas they are stronger along the light-mass direction in our heterostructures. Furthermore, we find that the type of substrate has a significant effect on the dispersion relations of the coupled modes. Also, by tuning the misalignment and separation between layers in double-layer phosphorene on polar substrates, we can engineer the hybrid modes. 

\end{abstract}

\maketitle
\date{\today}
\maketitle
\maketitle

\section{Introduction}

Phosphorene, a monolayer of black phosphorus (BP), has recently attracted special attention among two dimensional materials (2DMs) due to its unique highly anisotropic electronic and optical properties \cite{2053-1583-4-2-025064,Xia:nat13,liu:nano15,PhysRevB.92.081408,PhysRevLett.114.066803,Samira:drag2016,zhu2016black,PhysRevB.93.075408,Sa:nano15,liu:nano15,constantinescu2016multipurpose,2053-1583-4-2-025071,Samira:conductivity2017}.
BP is the most stable allotrope of phosphorus at room temperature and pressure. Few-layer phosphorene
can be obtained through mechanical exfoliation method akin to graphene \cite{C4CC05752J,hanlon2015liquid}. However, unlike graphene the phosphorene layers are not perfectly flat and form a puckered surface due to the sp$^{3}$ hybridization of 3s and 3p atomic orbitals. Also, the band gap of BP is direct and can be tuned from 0.3 eV to the visible part of the spectrum \cite{Rodin:prl2014,PhysRevB.90.205421}. In contrast, graphene is gapless, and the transition metal dichalcogenides (TMDs) have an indirect gap in the bulk phase and only monolayer TMDs have a direct gap\cite{Wang:nnano12}. Moreover, BP exhibits a strong in-plane anisotropy, which is absent in graphene and TMDs \cite{Phosphorene:review}.

In comparison to graphene, phosphorene is chemically reactive and tends to form strong bonds with surface of substrates which lead to some structural changes  \cite{C4CC05752J,2053-1591-3-2-025013,Vazifehshenas20104899}. Naturally, chemically stable 2DMs, such as graphene and BN, may be used for protecting fragile and low-chemical-stable 2DMs, such as phosphorene \cite{Bokdam:nanol11}.

The properties of substrate often drastically alter the transport behavior of the 2D crystal and the overall characteristics of the device. Recently, phosphorene has been transferred to the \BN\ substrate \cite{doganov2015transport,2053-1583-3-3-031007}. The interaction between  phosphorene and substrate is considered to play a crucial role in the modulation of the electronic properties of phosphorene-based devices\cite{gao2016critical,doganov2015transport}. In most currently available 2DMs, a sample lies on the top of a polar substrate such as \BN, SiO$_{2}$, SiC or Al$_{2}$O$_{3}$ \cite{dean2010boron,chen2015high,PhysRevB.93.085417,ishigami2007atomic,dai2015graphene,zarenia2012substrate,buscema2014effect}. In such heterostructures, the polar optical phonon modes of the substrate are localized near the 2DMs-substrate interface and are coupled to the surface optical (SO) phonon modes of the polar substrates through the long-range Fr$\ddot{\mathrm{o}}$hlich interaction. As a result, the SO phonon could be considered as the dominant plasmon-SO phonon coupling source in 2DMs on polar substrates.

\begin{figure}[ht]
	\includegraphics[width=8.8cm]{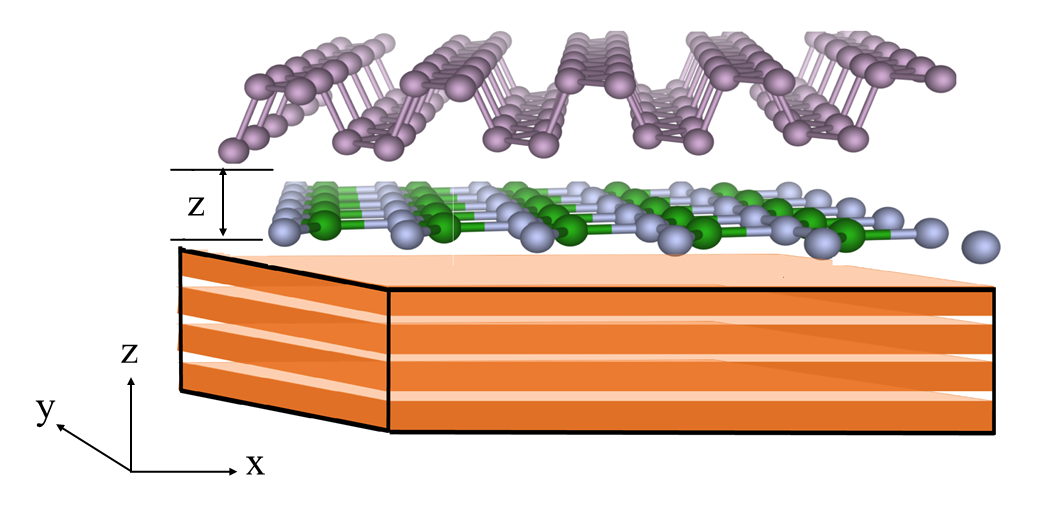}
	\caption{Schematic figure of a monolayer phosphorene where a polar insulating material (as an example \BN) has been used as a substrate. $z$ indicates the vertical distance of monolayer phosphorene from substrate.}
	\label{Fig1}
\end{figure}
The coupling of SO phonon modes to the plasma oscillations of free carriers is known as plasmon-SO phonon coupling. These coupled modes may be observed by the infrared (IR) transmission measurements\cite{yan2013damping,liu2016localized}. The plasmon-SO phonon coupling changes the dips in the IR reflectivity spectra from isolated plasmon and SO phonon frequencies to the normal coupled ones \cite{brar2014hybrid,stroscio2001phonons}.  

The coupled plasmon-SO phonon modes are extensively investigated in 2DMs with isotropic band structure such as graphene, silicene and germanene\cite{chen2008intrinsic,perebeinos2010inelastic,lin2013surface,wang2015energy}. It is shown that this phenomenon modifies many-body properties such as plasmon modes\cite{hwang2010plasmon,brar2014hybrid,xiaoguang1985plasmon} and self energy\cite{hwang2013surface} and can also change the mobility \cite{kaasbjerg2012phonon,ong2012theory}, effective mass\cite{jalabert:prb1989,hwang2013surface}, scattering rate \cite{perebeinos2008effects,lin2013surface} and inelastic lifetime \cite{ahn2014inelastic} in the carrier transport phenomena. Moreover, it can be considered as a mechanism for tuning the band gap \cite{PhysRevB.93.085417,fratini2008substrate,wang2015energy}.

In this paper, we model a system of phosphorene on polar insulator substrates as shown in Fig. \ref{Fig1}. We go beyond the assumption of independent electron and SO phonon modes and consider the coupled plasmon-SO phonon oscillations. Here, we theoretically study the coupled plasmon-SO phonon modes in monolayer and double-layer phosphorene on polar substrates using perturbation theory.
We start from the expression for dynamical dielectric function of the coupled system in the random phase approximation (RPA) level of theory, and develop a general formalism, which includes the effect of anisotropic energy dispersion and rotationally misaligned double-layer system. In such systems having the interaction between the electrons and substrate SO phonons could yield to phonon-mediated electron-electron interaction which creates a new set of collective modes with highly anisotropic dispersion. We find that in the case of phosphorene in contrast to graphene \cite{hwang2010plasmon,brar2014hybrid} and TMDs \cite{zhao2016plasmon}, the mode coupling effect not only modifies the plasma dispersion relation but also enables us to tune the hybrid plasmon-SO phonon modes in two crystallographic directions. So, the anisotropy is an important feature of the coupled electron-SO phonon oscillations spectrum in phosphorene systems. 

The paper is organized as follows. In Sec.\ref{theory} we present the theory  to calculate the generalized dielectric function in the presence of electron-SO phonon interaction. From the generalized dielectric function, we calculate the coupled plasmon-SO phonon modes in Sec.\ref{RESULTS AND DISCUSION} and present our results for the coupled plasmon-SO phonon modes in monolayer and double-layer phosphorene. 
Finally, we summarize the remarks of this work in Sec.\ref{Conclusion}.
\maketitle
\section{Theory} \label{theory}

 For an isolated phosphorene system, the anisotropic energy bands in the absence of electron-phonon interaction can be obtained from the $\mathbf{k}.\mathbf{p}$ free electron Hamiltonian: 

\begin{equation}
\begin{aligned}
H_{0}=
\begin{bmatrix}
E_{c}+\eta_{c}k_{x}^{2}+\nu_{c}k_{y}^{2} &\gamma k_{x}+\beta k_{y}^{2} \\
\gamma k_{x}+\beta k_{y}^{2}  & E_{\nu}-\eta_{\nu}k_{x}^{2}-\nu_{\nu}k_{y}^{2} 
\end{bmatrix} \ ,
\end{aligned}
\end{equation}
where E$_{c}$ (E$_{v}$) is the energy of 
conduction (valence) band edge and  the effective couplings between the bands are described by $\gamma$ and $\beta$
parameters. The $\eta_{c/v}$ and $\nu_{c/v}$ are related to the effective masses along $x$ and $y$ directions in the bands. These effective masses can be used to obtain an approximate anisotropic energy dispersion for phosphorene monolayer near the conduction (valence) band minimum (maximum)\cite{Rodin:prl2014}.   

In a phosphorene multilayer system with no electron-phonon interaction, the electrons in each layer interact with themselves and also with the electrons in other layers through the following electron-electron interaction Hamiltonian: 

\begin{equation}
H_{el-el}=\frac{1}{2} \sum\limits_{ij}\sum\limits_{\mathbf{kqp}} v_{ij}(q)  a^{\dagger}_{\mathbf{k+q},i} a^{\dagger}_{\mathbf{p-q},j} a_{\mathbf{p},j} a_{\mathbf{k},i} \ , 
\label{Hinter}
\end{equation}
where $v_{ij}(q)=v(q)e^{-qd_{ij}(1-\delta_{ij})}$ represents the diagonal (intralayer with $i=j$) and off-diagonal (interlayer with $i\neq j$) elements of the  bare Coulomb potential matrix. Here, we define $v(q)=2\pi e^2 / q \epsilon_{\infty}$ with $\epsilon_{\infty} $ being the high-frequency dielectric constant and $d_{ij}$ is the distance between \textit{i}-th and \textit{j}-th layers. Also, $a_{k,i}$ $(a_{k,i}^{\dagger})$ is the electron annihilation (creation) operator in layer $i$. 

If the phophorene layers are supported by polar materials, an additional interaction term involving the electron-SO phonon coupling can be included by using Fr$\ddot{\mathrm{o}}$hlich Hamiltonian  \cite{PhysRevLett.99.236802}: 
  
\begin{equation}
\begin{aligned}
H_{el-ph}=\sum\limits_{i}\sum\limits_{\lambda}\sum\limits_{\mathbf{kq}} [M_{0}^{\lambda}((q)] a^{\dagger}_{\mathbf{k+q},i} a_{\mathbf{k},i}(b_{\mathbf{q}\lambda}+b^{\dagger}_{\mathbf{-q}\lambda}) \ ,
\label{eq:mod3}
\end{aligned}
\end{equation} 
 which renormalizes the screened electron-electron potential through the dielectric function matrix, $\epsilon_{ij} (\omega,\mathbf{q})$:
 
 \begin{equation}
 V_{ij}^{sc}(\omega,\mathbf{q})=\frac{v_{ij}(q)}
 {det|\epsilon_{ij} (\omega,\mathbf{q})|} \ ,
 \label{Vsc}
 \end{equation}
 
 In Eq. (\ref{eq:mod3}), $b_{\mathbf{q}\lambda}$ ($b_{\mathbf{q}\lambda}^{\dagger}$) is the phonon annihilation (creation) operator with wave vector $\mathbf{q}$ and branch index $\lambda$ and  $M_{0}^{\lambda}(q)$ is the amplitude of the electron-SO phonon interaction \cite{Mahan:prb1972,hwang2010plasmon}:
\begin{equation}
\begin{aligned}
M_{0}^{\lambda}(q)=\big[v(q)\alpha \frac{\omega_{SO}^\lambda}{2} e^{-2qz}\big]^{1/2} \ ,
\label{eq4}
\end{aligned}
\end{equation}
where  $\omega_{SO}^\lambda$ is the SO phonon frequency of the $\lambda^{th} $  branch, $z$ is the vertical distance between phosphorene and substrate and we define:
  
\begin{equation}
\begin{aligned}
\alpha=\epsilon_\infty \big[\frac{1}{\epsilon_\infty+1}-\frac{1}{\epsilon_0+1}\big] \ .
\label{eq5}
\end{aligned}
\end{equation}
 with $\epsilon_o$ being the zero-frequency dielectric constant. Here, we assume that the phonon-phonon interaction is negligible so each mode couples to the electrons, independently. 
 
 In order to study the effect of electron-SO phonon coupling on the collective charge-density excitations, one needs to obtain zeros of the determinant of the dynamical dielectric function matrix (the poles of the screened potential). We use the dynamical RPA dielectric function in which the contribution of the electron-SO phonon interaction is taken into account\cite{zhang:jpcm1993}: 
\begin{equation}
\epsilon_{ij} (\omega,\mathbf{q})=\delta_{ij}-U_{ij}(\omega,q)\Pi_{i}(\omega,\mathbf{q}) \ ,
\label{eq8}
\end{equation}

Here, $U_{ij}(\omega,q)$ is the combined Coulomb and phonon-mediated interactions between $i$th and $j$th layers:
\begin{equation}
U_{ij}(\omega,q)=U_{0}(\omega,q)e^{-qd_{ij}(1-\delta_{ij})} \ ,
\label{eq9}
\end{equation}
where $U_{0}(\omega,q)=v_{ph}(\omega,q)+v(q)$ and the SO phonon-mediated electron-electron interaction, $v_{ph}(\omega,q)$, is given by \cite{sarma:Anphy1985}:
\begin{equation}
v_{ph}(\omega,q)=\sum_{\lambda}[M_{0}^{\lambda}(q)]^{2}D_{0}^{\lambda}(\omega) \ ,
\label{eq6}
\end{equation}
 $ D_{0}^{\lambda}(\omega)$ is the bare propagator for a phonon of branch index $\lambda$:
\begin{equation}
D_{0}^{\lambda}(\omega)= \frac{2\omega^{\lambda}_{SO}}{\omega^{2}-(\omega^{\lambda}_{SO})^{2}} \ .
\label{eq7}
\end{equation}
 Also, $\Pi_{j}(\omega,\mathbf{q})$ represents the noninteracting dynamic polarization  function of layer $j$ which is given by the following expression:
\begin{equation}
\Pi_{j}(\omega,\mathbf{q})=\frac{g_{s}}{\nu} \sum \limits_{\mathbf{k}} \frac{f^{j}(E_{\mathbf{q}})-f^{j}(E_{\mathbf{k+q}})}{E_{\mathbf{q}}-E_{\mathbf{k+q}}+\hbar\omega+i\eta} \ , 
\label{eq10}
\end{equation}
where $f^{j}(E_{\mathbf{q}})$ is the Fermi distribution function of j\textit{th} layer at energy $E$ corresponding to the wave vector $\mathbf{q}$, $g_{s}=2$ is spin degeneracy and $\eta$ is the broadening parameter which accounts for the disorder in the system. We use the zero-temperature polarization function because this approximation is valid at the typical doping densities $n=5 - 50\times10^{12}$ cm$^{-2}$ where the corresponding Fermi temperature ($T_{F}\approx 400- 800$K) for phosphorene is higher than room temperature. However, the zero-temperature dynamic polarization function for intraband transitions in an anisotropic 2D material can be calculated by making use of the following anisotropic parabolic energy dispersion relation
\begin{equation}
E_{\mathbf{k}}=\frac{\hbar^2}{2}(\frac{k_{x}^{2}}{m_{x}}+\frac{k_{y}^{2}}{m_{y}}) \ ,
\label{eq11}
\end{equation}
in Eq. (\ref{eq10}). So one obtains
\begin{equation}
\begin{aligned}
\frac{\Pi_{i}(\omega,\mathbf{q})}{g_{2d}}= (\frac{1}{Q_{i}})\Bigg[&\bigg(Z_{-}-sgn(\Re Z_{-})\sqrt{Z_{-}^{2}-1}\bigg)
  \\-&\bigg(Z_{+}-sgn(\Re Z_{+})\sqrt{Z_{+}^{2}-1}\bigg)\Bigg] \ .
\label{eq12}
\end{aligned}
\end{equation}    

 We define $\mathbf{Q}=\sqrt{m_{d}/\hat{M}}(\mathbf{q}/k_{F})$ and $\mathbf{K}=\sqrt{m_{d}/\hat{M}}(\mathbf{k}/k_{F})$ where $\hat{M}$ is the mass tensor with diagonal elements $m_{x}$ and $m_{y}$ along $x$ and $y$ direction, $m_{d}=\sqrt{m_{x}m_{y}}$, $g_{2d}=m_{d}/\pi \hbar^{2}$ and $Z_{\pm}=((\hbar\omega+i\eta)/\hbar Q k_{F} \nu_{F})\pm(Q/2)$  with $\nu_{F}=\hbar k_{F}/m_{d}$. By introducing the rotational angle, $\tau_{i}$, as the angle between $x$-axis in the laboratory frame and $x$ direction of the $i$th layer, we can write $Q_{i}(\theta)=q\sqrt{m_{d}R_{i}(\theta)}/k_{F}$  in which the orientation parameter, $R_{i}(\theta)$, defined as\cite{Samira:drag2016}:
\begin{gather}
R_{i}(\theta)=\bigg(\frac{\cos^{2}(\theta-\tau_{i})}{m_{x}}+\frac{\sin^{2}(\theta-\tau_{i})}{m_{y}}\bigg) \ .
\label{eq13}
\end{gather} 
In the case of monolayer, we will have $\tau_{i}=0$. The knowledge of the appropriate limiting behavior of the polarization function is important in investigating the excitation spectrum  specially the collective excitations of the system. Therefore, we obtain the polarization function in the dynamic long-wavelength limit where the plasmon excitation are important due to their long lifetimes. In this limit, the polarization function of Eq. (\ref{eq12}) can be approximated as (see the appendix \ref{App:AppendixA}) 
\begin{equation}
\frac{\Pi_{i}(\omega,q,\theta)}{g_{2d}}\approx R_{i}(\theta)E_{F}\frac{q^{2}}{\omega^{2}} \ .
\label{eq14}
\end{equation} 
In the following section, we present our calculations for the coupled plasmon-SO phonon modes in both monolayer and double-layer phosphorene systems with a number of experimentally chosen polar substrates/spacers. 
\begin{figure*}
\includegraphics[width=17.0cm]{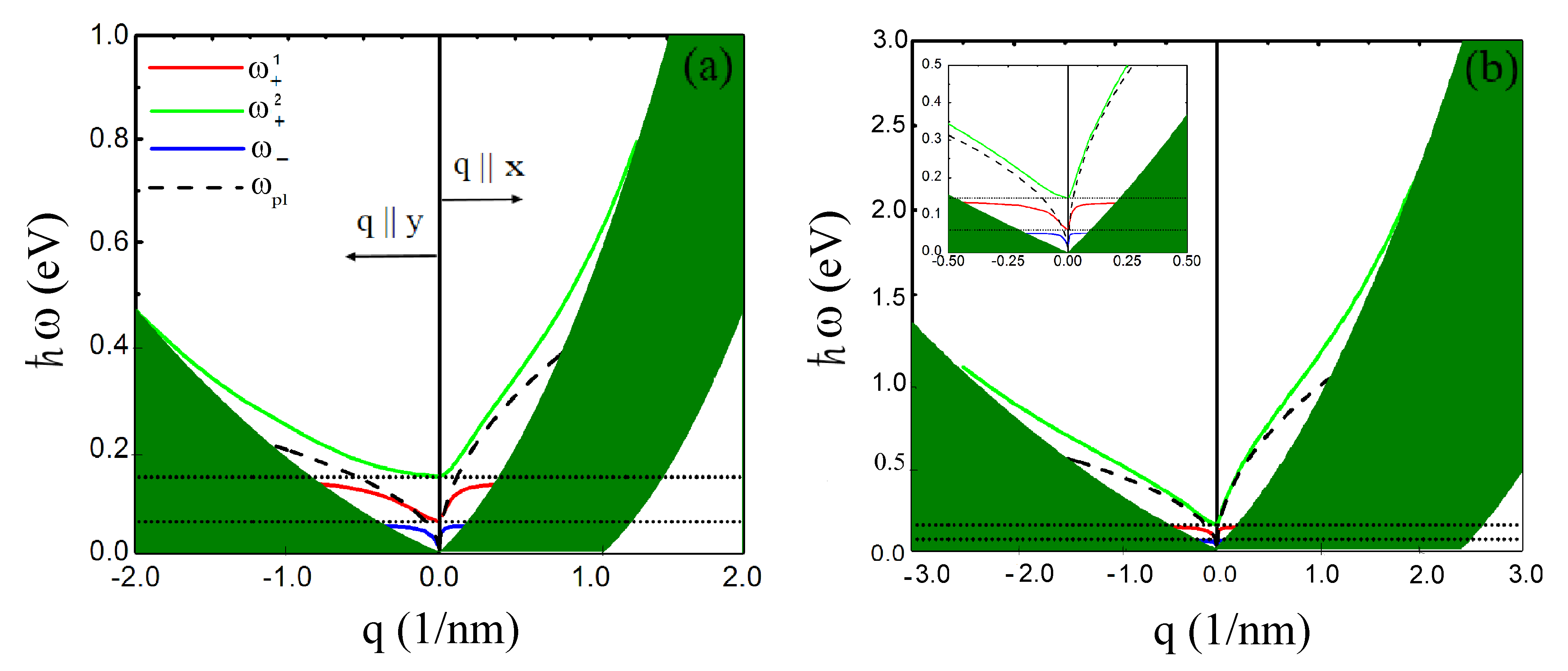}
\caption{The hybrid plasmon-SO phonon dispersions in phosphorene monolayer on $SiO_{2}$ substrate as a function of wave vector $q$ for two main crystallographic directions of phosphorene: $\theta=0$ $(q\parallel x)$ and $\theta=\pi/2$ $(q\parallel x)$ with (a) $n=1\times10^{13}$ cm$^{-2}$ (b) $n=5\times10^{13}$ cm$^{-2}$ and $z=0.2$ nm. The uncoupled plasmon dispersion is shown by dashed line. The two horizontal lines represent the frequencies of SO phonon modes, \textit{i.e.} $\hbar\omega_{so}^{1}=60$ meV and $\hbar\omega_{so}^{2}=146$ meV.}
\label{Fig2}
\end{figure*}
 
\maketitle
\section{ RESULTS AND DISCUSSION } \label{RESULTS AND DISCUSION}
\maketitle
\subsection{Coupled plasmon SO-phonon modes in monolayer phosphorene }
\label{sec:A}

 The RPA dielectric function for a monolayer system in which the electrons are coupled to SO phonons of a polar substrate can be obtained through summing over all the bare bubble diagrams as \cite{hwang2010plasmon} 

\begin{equation}
 \epsilon_{i} (\omega,\mathbf{q})=1-\frac{2\pi e^{2}}{\epsilon_{\infty}q} \Pi_{i}(\omega,\mathbf{q})+\sum_{\lambda}^{}\frac{\alpha e^{-2qz}}{1-\alpha e^{-2qz}- \omega^{2}/(\omega_{so}^{\lambda})^{2}} \ .
 \label{eq16}
\end{equation}
In the long-wavelength limit $(q\to 0)$ by inserting Eq. (\ref{eq14}) into Eq. (\ref{eq16}), we get the following coupled collective modes (see appendix \ref{App:AppendixB}):
\begin{equation}
\omega_{(+)}^{\lambda}(q,\theta)=\omega^{\lambda}_{so} \big(1+\alpha e^{-2qz}\frac{\omega^{2}_{pl}(q,\theta)}{(\omega_{so}^{\lambda})^{2}}\big) \ . 
\label{eq17}
\end{equation} 
\begin{equation}
\omega_{(-)}(q,\theta)=\omega_{pl}(q,\theta)\big(1-\frac{\alpha}{2} e^{-2qz}\big) \ .
\label{eq18}
\end{equation}  
\begin{figure}
\includegraphics[width=9.0cm]{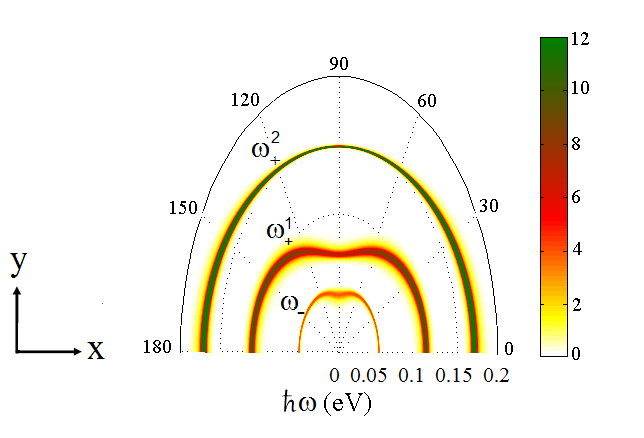}
\caption{Loss function, $|\Im(1/\det \epsilon(\omega,q,\theta))|$, for $q$=0.1 nm with n = 1$\times$10$^{13}$cm$^{-2}$ and $\eta$ = 5 meV. The radial and azimuthal coordinates are $\omega$ and the angular orientation of $\mathbf{q}$, respectively.}
 \label{Fig3}
\end{figure}  

 \begin{table}[hb]
 	\caption{Physical parameters of selected polar materials from Ref$\cite{PhysRevB.93.085417}$.}
 	\begin{ruledtabular}
 		\begin{tabular}{c c c c }
 			
 			& \qquad\quad SiO$_{2}$ &  \BN &  Al$_{2}$O$_{3}$\\ [0.5ex]  
 			
 			\hline 
 			
 			\\
 			$\hbar \omega^{1}_{SO}$(meV) &\qquad\quad 60 & 101 & 55 \\
 			\\
 			$\hbar \omega^{2}_{SO}$(meV) &\qquad\quad 146 & 195 & 94 \\
 			\\
 			$\epsilon_{0}$ &\qquad\quad 3.9 &5.1 &  12.5 \\
 			$\epsilon_{\infty}$ &\qquad\quad 2.5 & 4.1 & 3.2 \\
 			$\alpha$ &\qquad\quad 0.2 & 0.132 & 0.525 \\
 			
 		\end{tabular}
 		\label{table:nonlin}
 	\end{ruledtabular}
 \end{table} 

\begin{figure*}
\includegraphics[width=18.0cm]{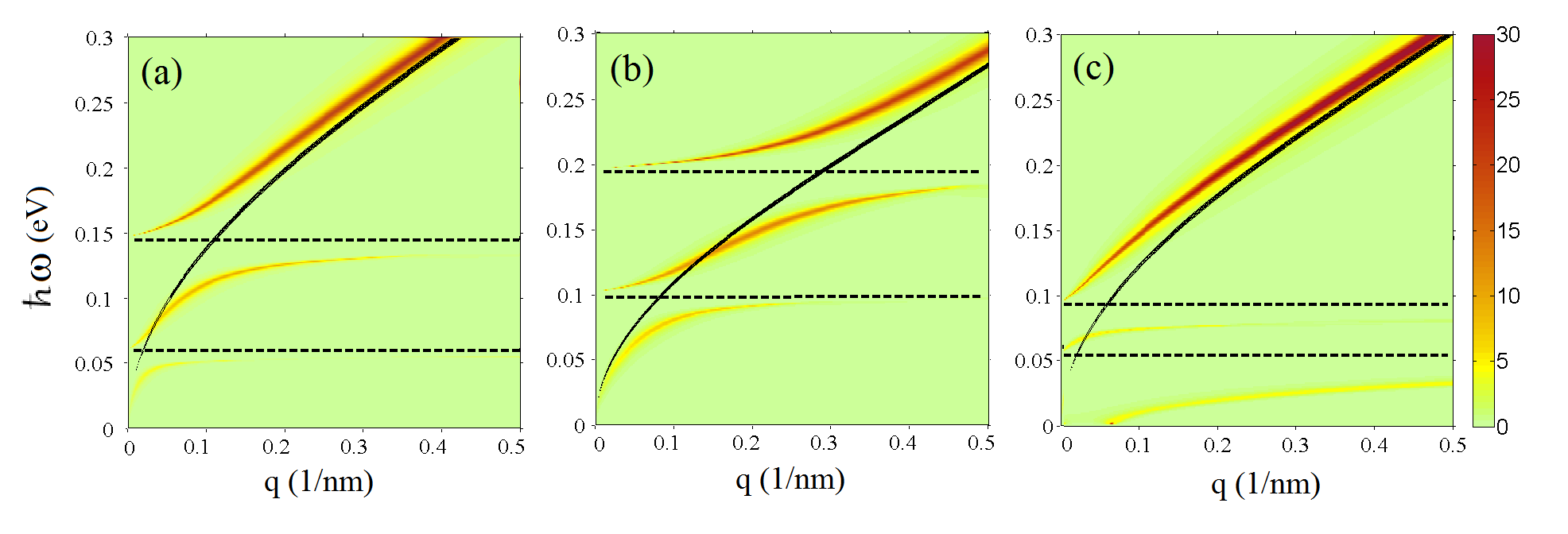}
\caption{(Color online) Effect of changing substrate: $|\Im(1/\det \epsilon(\omega,q,\theta))|$ for phosphorene monolayer at $\theta=0$ $(q\parallel x) $. The solid black line is the uncoupled plasmon branch and the two horizontal black dashed lines represent the frequencies of SO phonon modes of substrate: (a) SiO$_{2}$, (b) \BN\ and (c)  Al$_{2}$O$_{3}$ with $n=1\times10^{13}$ cm$^{-2}$, $z=$0.2 nm and $\eta$=5 meV.}
\label{Fig4}
\end{figure*}      
where $\omega_{pl}(q,\theta)=\sqrt{2\pi n e^{2}R_{i}(\theta)q/\epsilon_{\infty}}$ is the plasmon frequency of uncoupled anisotropic system. We consider a phosphorene monolayer sandwiched by polar materials such as SiO$_{2}$, \BN\ or Al$_{2}$O$_{3}$ which their physical parameters are listed in Table \ref{table:nonlin}. In addition, the effective masses in $x$ and $y$ directions of phosphorene monolayer are given as  $m_{x}\approx 0.15m_{0}$ and $m_{y}\approx 0.7m_{0}$ where $m_{0}$ is the free electron mass\cite{Low:prb14}.

In Fig. \ref{Fig2}, we depict the coupled and uncoupled plasmon modes of phosphorene monolayer on SiO$_{2}$ substrate calculated for two main crystallographic directions $\theta=0$ $(q\parallel x)$ and $\theta=\pi/2$ $(q\parallel y)$. According to this figure in the case of SiO$_{2}$ as the substrate, there are three hybrid modes, one plasmon-like, $\omega _{-}(q,\theta)$, and two phonon-like, $\omega^{\lambda} _{+}(q,\theta)$, as were previously predicted theoretically \cite{yan2013damping} and shown experimentally \cite{yan2013damping} for graphene  monolayer. It can be seen that the $\omega _{-}(q,\theta)$ (plasmon-like mode) is lower in energy than the uncoupled plasmon mode, $\omega_{pl}(q,\theta)$, and the lower (upper) phonon-like branch is starts from $\omega_{so}^{1} $ ($\omega_{so}^{2} $) at $q=0$ and increases below (above) the uncoupled plasmon dispersion by increasing $q$ along both directions. The coupling is strong when the frequencies of phonon-like modes deviate considerably from the bare SO phonon modes, $\omega_{so}^{\lambda}$ (see Eq. (\ref{eq17})).
As expected in 2DMs with anisotropic band structure, the hybrid plasmon-SO phonon modes have higher resonance frequencies\cite{Rodin:prb15,Low:prl14,Jin:prb15} and the plasmon-like mods are more affected by the electron-SO phonon coupling in the direction of the lighter mass, $q\parallel x$.
In the coupled system, the electrons can be scattered either by the emission of hybrid plasmon-SO phonon modes ($\omega_{+},\omega_{-}$) or by the single particle excitation (SPE). The boundary of SPE continuum represented by green shaded area in Fig. \ref{Fig2} is given by:
\begin{equation}
	\hbar \omega _{(SPE)}^{\pm}(q,\theta)=\frac{R_{i}(\theta)\hbar^{2}}{2}\pm q\nu_{F}\sqrt{R_{i}(\theta)m_{d}} \ .
	\label{eq19}
\end{equation} 

Though all hybrid plasmon-SO phonon excitations are damped at IR frequencies, the plasmon-like mode, $\omega_{-}$, experiences the Landau damping at smaller $q$ with respect to the phonon-like modes. On the other hand, the higher frequency phonon-like mode, $\omega^{2}_{+}$, stays away from $\omega^{2}_{SO}$ and enters the SPE region at large $q$ for both directions. Therefore, this mode shows a strong coupling and can be easily detected. 

Moreover, at high densities (Fig. \ref{Fig2}(b)) where the Fermi energy exceeds the SO phonon energy \textit{i.e.} $ E_{F}$ $\gg$ $\hbar \omega^{2}_{SO}$, the upper phonon-like branch strongly deviates from the uncoupled SO phonon's energy while $\omega^{1}_{+}$ branch still remains between the two phonon energies. Thus, the high energy phonon-like mode, $\omega_{+}^{2}$, which is very sensitive to the density and direction, can be considered as a tunable quantity for applications in optical plasmonic devices.

Furthermore, we study the impact of angular orientation of $\mathbf{q}$ on the behavior of hybrid plasmon-SO phonon modes by plotting the loss function, $|\Im(1/\det \epsilon(\omega,q,\theta))|$, for $q$=0.1 nm (see Fig. \ref{Fig3}). The results show that the maximum value of loss function  occurs around  $\theta$= 0 and 180 (along $x$ direction) for all three branches.
 
In order to explore the effects of specific substrate on the hybrid plasmon-SO phonon modes, we show the loss functions of phosphorene monolayer on (a) SiO$_{2}$, (b) \BN\ and (c) Al$_{2}$O$_{3}$ polar substrates along $x$ direction ($\theta=0$) in Fig. \ref{Fig4}. One can see that using the Al$_{2}$O$_{3}$ (with higher $\alpha$) as a polar substrate results in the strong coupling as evidenced by considerable deviation of $\omega_{-}$ from $\omega_{pl}$ and of $\omega_{+}^{\lambda}$ from $\omega_{SO}$. 

It sould be pointed out that while the phonon frequency $\omega_{SO}$ is an important parameter for the phonon-like modes, $\omega_{+}^{\lambda}$, the $\alpha$ parameter mostly affects the plasmon-like spectrum. Hence, the choice of substrate can be used to engineer the plasmon dispersion in phosphorene. 
 
\subsection{Coupled plasmon-SO phonon modes in double-layer phosphorene}
\label{sec:B}             	   

\begin{figure}
\includegraphics[width=9.5cm]{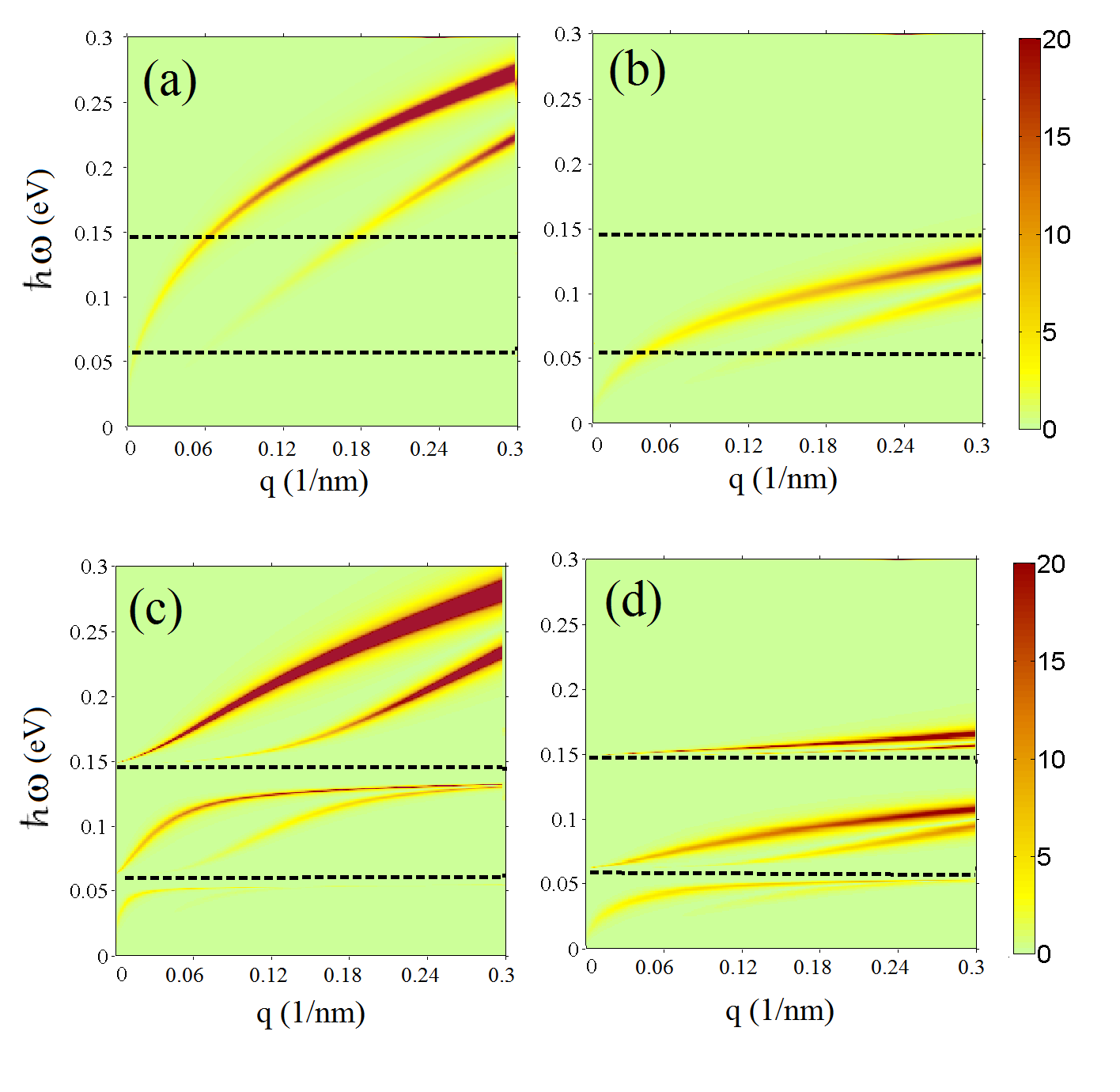}
\caption {\color{black}(Color online) (a) , (b) The uncoupled and (c) and (d) coupled plasmon-SO phonon modes for two aligned phosphorene monolayers sandwiched by SiO$_{2}$ as a function of wave vector $q$ for two main crystallographic directions: (a) and (c) $\theta=0$ $(q\parallel x)$ and (b) and (d) $\theta=\pi/2$ $(q\parallel y)$ with $n=1\times10^{13}$ cm$^{-2}$, d$_{12}$=5 nm $z$=0.2 nm and $\eta$=5 meV.} 
\label{Fig5}
\end{figure}

Here, we consider a double-layer phosphorene with equal electron densities sandwiched by a homogeneous dielectric medium that models the substrate. It is reasonable to calculate the uncoupled plasmon modes before discussing the hybrid modes from the zeros of determinant of dielectric function matrix Eq. (\ref{eq8}).  
In the leading-$q$ approximation (long-wavelength limit), two plasmonic branches are obtained through the following relations\cite{Rodin:prb15}:  
\begin{equation}
 \omega_{ac}(q,\theta)=2q\sqrt{\frac{n\pi e^2d_{12}}{\epsilon_{\infty}}\frac{R_{1}(\theta)R_{2}(\theta)}{R_{1}(\theta)+R_{2}(\theta)}} \ .
 \label{eq20}
\end{equation}
\begin{equation}
 \omega_{op}(q,\theta)=\sqrt{\frac{2n\pi e^2q}{\epsilon_{\infty}}\big(R_{1}(\theta)+R_{2}(\theta)\big)} \ .
 \label{eq21}
\end{equation}
{\color{black} These two uncoupled branches are shown in Fig. \ref{Fig5}(a) for $\theta=0$ (along $x$ direction) and in Fig. \ref{Fig5}(b) for $\theta=\pi/2$ (along $y$ direction) when SiO$_{2}$ is considered as a substrate/spacer \cite{Rudenko:prb14,Rodin:prb15}. 
	
It can be seen that the plasmon modes experience a stronger reduction along $y$ direction compared to $x$ direction.  
For the case of electron-SO phonon coupling in a double-layer phosphorene, we find two acoustic phonon-like modes, $\omega^{\lambda}_{ac(+)}$, with $\lambda$=1,2 and one acoustic plasmon-like mode, $\omega_{ac(-)}$ (see appendix \ref{App:AppendixB}). The long-wavelength dispersions of the acoustic modes can be written as}:
\begin{equation}
\omega^{\lambda}_{ac(+)}(q,\theta)=\omega^{\lambda}_{SO}\sqrt{1+\frac{4n\pi d_{12} e^2q^{2}\alpha e^{-2qz}}{(\omega^{\lambda}_{SO})^{2}\epsilon_{\infty}}\frac{R_{1}(\theta)R_{2}(\theta)}{R_{1}(\theta)+R_{2}(\theta)}} \ .
\label{eq22}
\end{equation}
\
 \begin{equation}
 \omega_{ac(-)}(q,\theta)=2q\sqrt{\frac{n\pi e^2d_{12}(1-\alpha e^{-2qz})}{\epsilon_{\infty}}\frac{R_{1}(\theta)R_{2}(\theta)}{R_{1}(\theta)+R_{2}(\theta)}} \ .
 \label{eq23}
 \end{equation}
 
In the same vein, we obtain two optical phonon-like modes, $\omega^{\lambda}_{op(+)}$, and one optical plasmon-like mode, $\omega_ {op(-)}$ as
\begin{equation}
\omega^{\lambda}_{op(+)}(q,\theta)=\omega^{\lambda}_{SO}\sqrt{1+\frac{2n\pi e^2q \alpha e^{-2qz}}{(\omega^{\lambda}_{SO})^{2}\epsilon_{\infty}}\big(R_{1}(\theta)+R_{2}(\theta)\big)} \ .
\label{eq24}
\end{equation}
\begin{equation}
\omega_ {op(-)}(q,\theta)=\sqrt{\frac{2n\pi e^2q(1-\alpha e^{-2qz})}{\epsilon_{\infty}}\big(R_{1}(\theta)+R_{2}(\theta)\big)} \ .
\label{eq25}
\end{equation}

\begin{figure*}
\includegraphics[width=18.5cm]{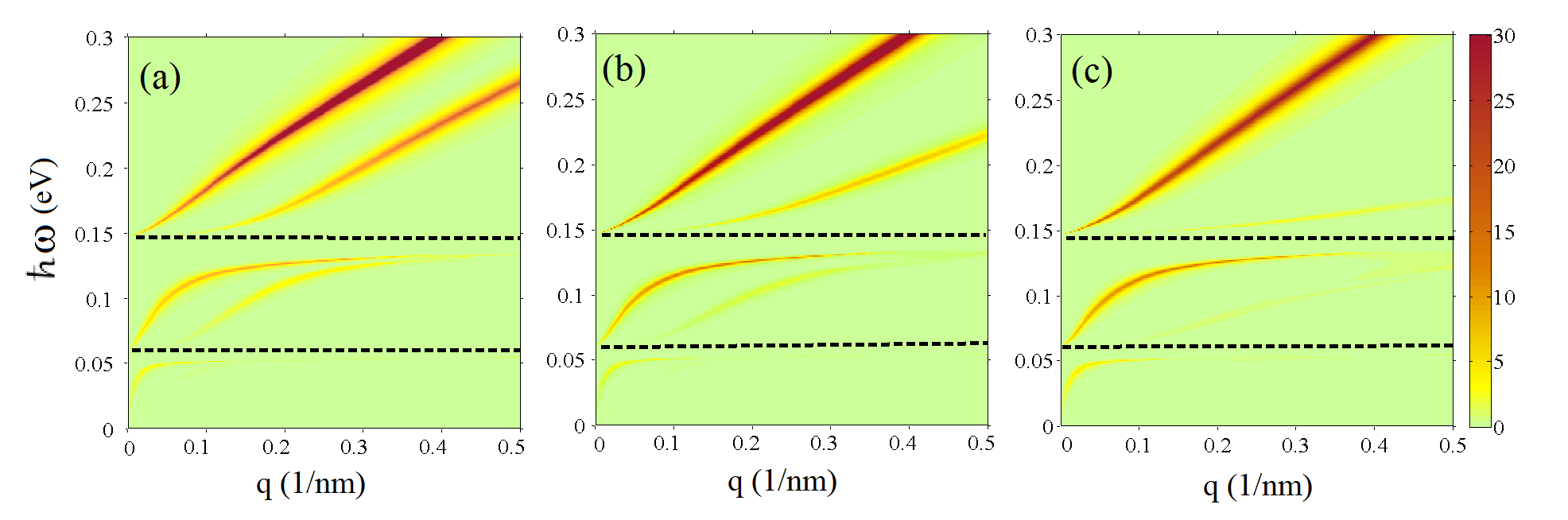}
\caption {(Color online) Effect of changing rotation angle on the loss function, $|-Im[1/\varepsilon(\omega,q)]|$, for double-layer phosphorene sandwiched by SiO$_{2}$ in $(\omega,q)$ space along $\theta=0$ $(q\parallel x)$ for $\tau_{1}=0$ and (a) $\tau_{2}=\pi/4$, (b) $\tau_{2}=\pi/3$ and (c) $\tau_{2}=\pi/2$ with d$_{12}=5$ nm, $n=1\times10^{13}$ cm$^{-2}$ and $z$=0.2 nm.}
\label{Fig6}
\end{figure*}
\begin{figure*}
 \includegraphics[width=18.5cm]{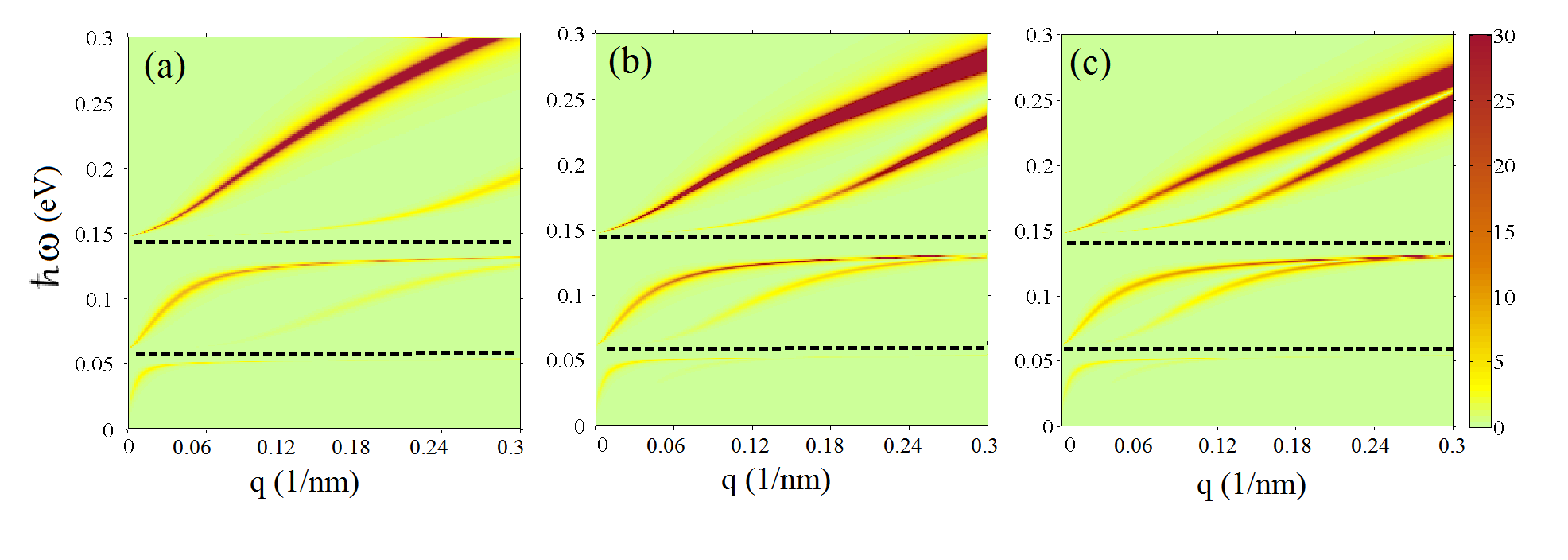}
 \caption {(Color online) Effect of changing separation between layers on the loss function, $|-Im[1/\varepsilon(\omega,q)]|$, for the two parallel phosphorene monolayers sandwiched by SiO$_{2}$ in $(\omega,q)$ space along $\theta=0$ $ (q\parallel x)$ for (a) $d_{12}=2$ nm (b) $d_{12}$=5 nm and (c) $d_{12}$=8 nm with $n=1\times10^{13}$ cm$^{-2}$ and $z$=0.2 nm.}
\label{Fig7}
\end{figure*} 
The behavior of the coupled modes with SiO$_{2}$ as substrate/spacer is presented in Fig. \ref{Fig5}(c) and \ref{Fig5}(d) for $\theta=0$ and $\theta=\pi/2$, respectively. The dispersion relations of $\omega_{op/ac(-)}$ modes resemble the uncoupled acoustic and optical plasmon modes with an additional  $(1-\alpha e^{-2qz})^{1/2}$ multiplier which makes these coupled modes lie lower than the uncoupled ones.
 {\color{black}  

  Similar to the case of phosphorene monolayer, the hybrid plasmon-SO phonon frequencies along $q\parallel x$ are larger than along $q \parallel y$ because the carriers along $y$ direction have larger mass and so get damped faster. One can also see that the uncoupled plasmon modes for $q \parallel y$ lie lower than $\omega_{+}^{2}$ and therefore plasmon-SO phonon coupling along $y$ direction is substantially weaker compared to the $x$ direction.

In order to understand how the orientation parameter, $R_{i}(\theta)$, impacts the behavior of hybrid plasmon-SO phonon modes in the double-layer system, we show three cases in which $\tau_{1}=0$ and $\tau_{2}$ progressively increases from $\pi/4$ to  $\pi/3$ and to  $\pi/2$ along $x$ direction and for substrate/spacer SiO$_{2}$ in Fig. (\ref{Fig6}). One may notice that, in general, with the reduction of $\tau_{2}$, the coupled plasmon-SO phonon modes become significantly stronger because the hybrid modes are larger along the lower mass direction, \textit{i.e.} $\theta=0$. Furthermore, it is clear that the three acoustic plasmon/phonon-like modes are very sensitive to the rotation of layers and get damped as the angle of rotation is increased.   
Finally, we address the effect of separation between layers, $d_{12}$, on the coupled modes along $x$ direction in Fig. (\ref{Fig7}) for the substrate/spacer SiO$_{2}$. Here,  increasing the separation between layers  shows a similar effect on hybrid mode frequencies as the reduction of rotation angle $\tau_{2}$ (see Fig. \ref{Fig6}). As expected, the hybrid acoustic branches, similar to the uncoupled ones, depend strongly on the separation between layers at long-wavelengths and move to the optical branches by increasing the layers' spacing.  
As a result, by adjusting two parameters $\tau_{2}$ and $d_{12}$, the acoustic branches may get strongly damped 
or a transition to the optical modes may be observed. } 

\section{Conclusion}  \label{Conclusion}

In summary, we have considered monolayer and double-layer phosphorene systems located on commonly used polar substrates (SiO$_{2}$, \BN\ and Al$_{2}$O$_{3}$) and calculated the anisotropic coupled plasmon-SO phonon dispersion relations. In the presented theory, the dynamical dielectric function is calculated within the RPA which includes the many-body electron-electron interaction in phosphorene layer(s) as well as the interaction between electrons and the long-range electric field generated by the substrate SO phonons. More importantly, we have obtained some analytical expressions for the plasmon-SO phonon dispersion relations in the long-wavelength limit which yield the same results of loss function from dielectric function.
In the long-wavelength limit, three hybrid plasmon-SO phonon branches are obtained due to two relevant SO phonon modes of polar substrates in monolayer phosphorene. In case of double-layer, these hybrid modes are doubled with three acoustic, $\omega_{ac(\pm)}$, and three optical modes, $\omega_{op(\pm)}$. We have shown that these hybrid modes are considerably stronger along  x direction because of lower effective mass of electrons. Moreover, by increasing the electron density, the hybrid excitation modes become stronger at both directions, simultaneously. 
Among mentioned substrates, we have found that Al$_{2}$O$_{3}$ as a polar substrate leads to more pronounced coupling of plasmon-phonon modes in comparison to \BN\ and SiO$_{2}$ for the electron density used here. 
Therefore, the phonon frequency of a substrate, $\omega_{SO}$, is the most effective parameter in phonon-like modes, $\omega^{\lambda}_{+}$, whereas $\alpha$ parameter changes the plasmon-like mode, $\omega_{-}$. Hence, the choice of substrate can be utilized in order to engineer the plasmon-SO phonon dispersion in phosphorene.
We have also investigated the effect of misalignment of two layers on the hybrid modes in double-layer phosphorene and found that the acoustic phonon-like modes are more affected by rotation. In addition, we have observed that by decreasing separation between two layers the acoustic modes move away from the optical modes and quickly damped. As a result, the rotation angle and separation between two layers can be used as a mechanism for tuning the plasmon-SO phonon coupling effects.

 \appendix
\section{\\The polarization function in long-wavelength limit} \label{App:AppendixA}     
The dynamic polarization function given by Eq. (\ref{eq10}) has the following expression in the long-wavelength limit at zero-temperature:
\begin{equation}  
 \begin{aligned}
 \frac{\Pi(\omega,q,\theta)}{g_{2d}}= \frac{1}{2}\bigg[-2&+\sqrt{(1-\frac{2\omega}{k_{F}\nu_{F} Q^{2}})^{2}-\frac{4}{Q^{2}}}
 \\
&+\sqrt{(1+\frac{2\omega}{k_{F}\nu_{F}Q^{2}})^{2}-\frac{4}{Q^{2}}}\quad \bigg] 
\label{pol1}
 \end{aligned}		
\end{equation}
One can rewrite Eq. (\ref{pol1}) as:
\begin{equation} 
  \begin{aligned}
  \frac{\Pi(\omega,q,\theta)}{g_{2d}}= \frac{1}{2}\bigg[-2&+(1-\frac{2\omega}{k_{F}\nu_{F} Q^{2}})\big(1-\frac{2/Q^{2}}{(1-\frac{2\omega}{k_{F}\nu_{F} Q^{2}})^{2}}\big)
  \\
  &+(1+\frac{2\omega}{k_{F}\nu_{F} Q^{2}})\big(1-\frac{2/Q^{2}}{(1+\frac{2\omega}{k_{F}\nu_{F} Q^{2}})^{2}}\big)\bigg]
  \label{pol2}
  \end{aligned}		
\end{equation} 
and find the approximate the relation:
\begin{equation} 
    \begin{aligned}
    \frac{\Pi(\omega,q \to 0,\theta)}{g_{2d}}\approx \frac{-2}{Q^2} \bigg(\frac{1}{1-\frac{4\omega^{2}}{(k_{F}\nu_{F}Q^{2})^{2}}}\bigg)
    \label{pol3}
    \end{aligned}		
\end{equation}
Finally, by ignoring unity in the denominator and making use of $Q_{i}(\theta)=q\sqrt{m_{d}R_{i}(\theta)}/k_{F}$ and $E_{F}=\hbar ^{2}k_{F}^{2}/2m_{d}$, one obtains Eq. (\ref{eq14}).         
\section{\\Electron-SO phonon modes in long-wavelength limit} \label{App:AppendixB}
     
 \subsection{Monolayer}
Inserting Eq. (\ref{eq14}) into Eq. (\ref{eq16}), we have: 
\begin{equation}
       \epsilon (\omega,q,\theta)=1-\frac{\omega_{pl}^{2}(q,\theta)}{\omega^{2}(q,\theta)} +\sum_{\lambda}^{}\frac{\alpha e^{-2qz}}{1-\alpha e^{-2qz}- \omega^{2}(q,\theta)/(\omega_{so}^{\lambda})^{2}}
       \label{eq1b}
\end{equation}
From the zeros of dielectric function (here we drop the $q$ and $\theta$ for simplicity): 
\begin{equation}
     \omega^{2}( \omega^{2}-(\omega_{SO}^{\lambda})^{2})-\omega_{pl}^{2}(\omega^{2}-(\omega_{SO}^{\lambda})^{2})-\omega_{pl}^{2}(\omega_{SO}^{\lambda})^{2} \alpha e^{-2qz}=0
      \label{eq2b}      
\end{equation}
the coupled plasmon-SO phonon modes are obtained as:
\begin{equation} 
         \begin{split}
          (\omega_{(\pm)}^{\lambda})^{2}=\frac{1}{2} \bigg[\big( (\omega_{SO}^{\lambda})^{2}+\omega_{pl}^{2}\big)&\pm \big[((\omega_{SO}^{\lambda})^{2}-\omega_{pl}^{2})\\
          &+4\omega_{pl}^{2}(\omega_{SO}^{\lambda})^{2} \alpha e^{-2qz}\big]^{1/2}\bigg]
         \label{B3}
         \end{split}
\end{equation}    
By Taylor expanding the right hand side of Eq. (\ref{B3}), we get:
\begin{equation}
\omega_{(+)}^{\lambda}(q,\theta)=\omega^{\lambda}_{so} \big(1+\alpha e^{-2qz}\frac{\omega^{2}_{pl}}{(\omega_{so}^{\lambda})^{2}-\omega_{pl}^{2}})
\label{eq4b}
\end{equation} 
\begin{equation}
\omega_{(-)}^{\lambda}(q,\theta)=\omega_{pl} \big(1-\frac{\alpha e^{-2qz}}{2} \frac{(\omega_{so}^{\lambda})^{2}}{(\omega_{so}^{\lambda})^{2}-\omega_{pl}^{2}}\big) 
\label{eq5b}
\end{equation}
Since $\omega_{pl} \to 0$ at the long-wavelength limit, one can safely ignore $\omega_{pl}$ in the denominator of the above equations and obtain Eqs. (\ref{eq17}) and (\ref{eq18}).

\subsection{Double-layer}
Here, we calculate the coupled plasmon-SO phonon modes in double-layer systems with anisotropic band structure. In such systems, the rotation of one layer with respect to the other should be considered. In order to determine the dispersion relation of the coupled modes, we need to calculate the zeros of  determinant of the total dielectric matrix (Eq. (\ref{eq8})) in the long-wavelength limit:
\begin{equation}
       \begin{split}
       \epsilon(\omega,q,\theta)=&1-(R_{1}(\theta)+R_{2}(\theta))\Pi^{\prime}(q,\omega)U_{0}(q,\omega)\\
       &+(U_{0}(q,\omega) \Pi^{\prime}(q,\omega))^{2} R_{1}(\theta)R_{2}(\theta) (1-e^{-2qd_{12}})
       \label{B6}
       \end{split}
\end{equation}
where $\Pi^{\prime}(\omega,q,\theta)=\Pi_{i}(\omega,q,\theta)/R_{i}(\theta)$.
By doing some algebra, we get the following relation: 
\begin{equation}
       \begin{split}
       \epsilon(\omega,q,\theta)&=U_{0}^{2}(q,\omega) (2qd_{12}) R_{1}(\theta)R_{2}(\theta) [\Pi^{\prime}(\omega,q,\theta)-\Pi_{+}(\omega,q,\theta)]
       \\
       &\times [\Pi^{\prime}(\omega,q,\theta)-\Pi_{-}(\omega,q,\theta)]
        \label{B7}
       \end{split}
\end{equation}
where $\Pi_{+}$ and $\Pi_{-}$ defined as 
\begin{equation}
        \Pi_{+}(\omega,q,\theta)=\frac{R_{1}(\theta)+R_{2}(\theta)}{(R_{1}(\theta)R_{2}(\theta)U_{0}(q,\omega)2qd_{12}}
        \label{B8}
\end{equation} 
and
\begin{equation}
                 \Pi_{-}(\omega,q,\theta)=\frac{1}{(R_{1}(\theta)+R_{2}(\theta))U_{0}(q,\omega)}
                 \label{eq9b}
\end{equation} 
The dispersion relation of the couple modes is given by $\Pi^{\prime}(\omega,q,\theta)=\Pi_{\pm}(\omega,q,\theta)$. Using the Eqs. (\ref{eq4}-\ref{eq7}) in the $\Pi^{\prime}(\omega,q,\theta)=\Pi_{+}(\omega,q,\theta)$ condition, we get: 
\begin{equation}
\begin{split}
             \omega^{2}\big( \omega^{2}   -(\omega_{SO}^{\lambda})^{2}\big)&-q^{2}f_{ac}(\theta)\big[\omega^{2}\\
             &-(\omega_{SO}^{\lambda})^{2}+(\omega_{SO}^{\lambda})^{2}\alpha e^{-2qz}\big]=0
\label{B10} 
\end{split}   
\end{equation}
 where $f_{ac}(\theta)$ is defined as:
\begin{equation}
             f_{ac}(\theta)=\frac{4\pi n e^{2}d_{12}}{\epsilon_{\infty}} \frac{R_{1}(\theta)R_{2}(\theta)}{R_{1}(\theta)+R_{2}(\theta)}
             \label{B11}
\end{equation}
After some algebra, we derive the following relation:
\begin{widetext}
\begin{equation} 
           (\omega_{ac({\pm})}^{\lambda})^{2}=\frac{ (\omega_{so}^{\lambda})^{2}+q^{2}f_{ac}(\theta)}{2} \pm \sqrt{\frac{\big((\omega_{SO}^{\lambda})^{2}-q^{2}f_{ac}(\theta)\big)^{2}+4q^{2}f_{ac}(\theta)(\omega_{SO}^{\lambda})^{2} \alpha e^{-2qz}}{4}}
           \label{B12}
\end{equation}
\end{widetext}
which can be simplified in the long-wavelength limit as:      
\begin{equation} 
            (\omega_{ac(+)}^{\lambda})^{2}=(\omega_{so}^{\lambda})^{2} \bigg[1 +\frac{f_{ac}(\theta)q^{2}\alpha e^{-2qz}}{(\omega_{so}^{\lambda})^{2}}\bigg]
            \label{B13}
\end{equation}
and      
\begin{equation} 
           (\omega_{ac(-)}^{\lambda})^{2}=f_{ac}(\theta)q^{2}[1-\alpha e^{-2qz}]
             \label{B14}
\end{equation}
Substituting the relation for $f_{ac}(\theta)$ in above relations, one obtains Eqs. (\ref{eq22}) and (\ref{eq23}) for the coupled acoustic modes. In the case of the coupled optical modes, we use the $\Pi^{\prime}(q,\omega)=\Pi_{-}(q,\omega)$ condition and get the following equation:  
\begin{equation}
     \begin{split}
     \omega^{2}\big( \omega^{2}-(\omega_{so}^{\lambda})^{2}\big)&-f_{op}(\theta)\big[\omega^{2}
     \\&-(\omega_{so}^{\lambda})^{2}+(\omega_{so}^{\lambda})^{2}\alpha e^{-2qz}\big]=0
     \label{B15}
     \end{split}      
\end{equation} 
where  $f_{op}(\theta)$ is given by: 
\begin{equation}
     f_{op}(\theta)=\frac{2\pi n e^{2}}{\epsilon_{\infty}} \big(R_{1}(\theta)+R_{2}(\theta)\big)
     \label{B16}
\end{equation}
Solving Eq. (\ref{B15}) yields:
\begin{widetext}
     \begin{equation} 
     (\omega_{op(\pm)}^{\lambda})^{2}=\frac{(\omega_{so}^{\lambda})^{2}+qf_{op}(\theta)}{2}
     \pm \sqrt{\frac{\big((\omega_{so}^{\lambda})^{2}-q f_{op}(\theta)\big)^{2}-4 qf_{op}(\theta)(\omega_{so}^{\lambda})^{2}(1-\alpha e^{-2qz})}{4}}
     \label{B17}
     \end{equation}
\end{widetext}
In the long-wavelength limit, we end up with the following expressions for the coupled optical modes:
\begin{equation} 
     (\omega_{op(+)}^{\lambda})^{2}=(\omega_{so}^{\lambda})^{2} \bigg[1 +\frac{qf_{op}(\theta)\alpha e^{-2qz}}{(\omega_{so}^{\lambda})^{2}}\bigg]
     \label{B18}
\end{equation}
and 
\begin{equation} 
     (\omega_{op(-)}^{\lambda})^{2}=qf_{op}(\theta)[1-\alpha e^{-2qz}]
     \label{B19}
\end{equation}
Finally, by inserting $f_{op}(\theta)$ into Eqs. (\ref{B18}) and (\ref{B19}) one obtains  Eqs. (\ref{eq24}) and (\ref{eq25}).  
      

%

\end{document}